\pdfoutput=1
\documentclass[10pt, conference, compsocconf]{IEEEtran}
\IEEEoverridecommandlockouts
\usepackage{subfigure}

\ifCLASSINFOpdf
   \usepackage[pdftex]{graphicx}
 
   \graphicspath{{../pdf/}{../jpeg/}}

  \DeclareGraphicsExtensions{.pdf,.jpeg,.png}
\else

  \usepackage[dvips]{graphicx}

  \graphicspath{{../eps/}}

  \DeclareGraphicsExtensions{.eps}
\fi

\usepackage[cmex10]{amsmath}
\usepackage{booktabs}
\usepackage{amsmath}
\usepackage{amssymb}
\usepackage{algorithm}
\usepackage{algorithmic}
\usepackage[hidelinks]{hyperref}
\usepackage{marvosym}
\renewcommand{\algorithmicrequire}{\textbf{Input:}}
\renewcommand{\algorithmicensure}{\textbf{Output:}}
\usepackage{url}
\usepackage{multicol}
\usepackage{ifsym}

\urldef{\mailsa}\path|{bing.huang,kwokyan.lam }@ntu.edu.sg|
\urldef{\mailsb}\path|boyuzhang@swin.edu.au|
\urldef{\mailsc}\path|michael.sheng@mq.edu.au|

\usepackage{geometry}
\begin{document}
\title{ Proactive Detection of Physical Inter-rule Vulnerabilities in IoT Services Using a Deep Learning Approach }

\author{
    \IEEEauthorblockN{Bing Huang$^{1}$, Chen Chen$^{1}$\Letter, Kwok-Yan Lam$^{1}$, Fuqun Huang$^{2}$}
    \IEEEauthorblockA{$^1$ School of Computer Science and Engineering, Nanyang Technological University, Singapore }
    \IEEEauthorblockA{$^2$ Department of Computer Science, Western Washington University, WA, United States}
    \IEEEauthorblockA{\{\href{mailto:bing.huang@ntu.edu.sg}{bing.huang}, \href{mailto:chen.chen@ntu.edu.sg}{chen.chen}, \href{mailto:kwokyan.lam@ntu.edu.sg}{kwokyan.lam}\}@ntu.edu.sg, \href{mailto:huangf2@wwu.edu}{huangf2}@wwu.edu }
    }

\maketitle

\begin{abstract}
Emerging Internet of Things (IoT) platforms provide sophisticated capabilities to automate IoT services by enabling occupants to create trigger-action rules. Multiple trigger-action rules can physically interact with each other via shared environment channels, such as temperature, humidity, and illumination. We refer to inter-rule interactions via shared environment channels as a \emph{physical inter-rule vulnerability}. Such vulnerability can be exploited by attackers to launch attacks against IoT systems. We propose a new framework to proactively discover possible physical inter-rule interactions from user requirement specifications (i.e., descriptions) using a deep learning approach. Specifically, we utilize the Transformer model to generate trigger-action rules from their associated descriptions. We discover two types of physical inter-rule vulnerabilities and determine associated environment channels using natural language processing (NLP) tools. Given the extracted trigger-action rules and associated environment channels, an approach is proposed to identify hidden physical inter-rule vulnerabilities among them.  Our experiment on 27983 IFTTT style rules shows that the Transformer can successfully extract trigger-action rules from descriptions with 95.22\% accuracy.  We also validate the effectiveness of our approach on 60 SmartThings official IoT apps and discover 99 possible physical inter-rule vulnerabilities. 
 
\end{abstract}


\section{Introduction}
The Internet of Things (IoT) is growing rapidly and reshaping our lifestyles. The number of IoT devices in use 
%is expected 
has grown 16\%, reaching more than 16.7 billion in 2023 \cite{2023IoT}. These IoT devices are heterogeneous in terms of communication protocols (e.g., Bluetooth, ZigBee, Z-Wave, and Wi-Fi) and programming language. Service-oriented Computing (SOC) is a promising paradigm for abstracting IoT devices as \emph{IoT services} by hiding the low-level implementation details of the protocols for communicating to the devices  \cite{issarny2016revisiting}. 
% An IoT service has functional and non-functional properties. For example, a light has the functionalities of turning on and turning off. Its non-functional properties may include its price. 
An IoT service includes both functional properties, which define its capabilities such as turning on and off, and non-functional properties, such as pricing information.

An important application domain of IoT is the smart home which is equipped with a myriad of IoT devices and services. The ultimate goal of smart homes is to improve people’s life quality by making their daily life more convenient, efficient, secure, and comfortable \cite{huang2021enabling}.  Customizing the smart home environment and automating IoT services by the Trigger-Action Programming (\textbf{TAP}) paradigm is one of the promising ways to achieve such goals. Occupants can compose multiple IoT services into an \emph{IoT rule} (a.k.a \emph{trigger-action rule}) in the form of ``IF triggers
happen, THEN perform an action'' \cite{corno2019recrules}. In this paper,  a composite IoT service is defined as a trigger-action rule.  An example of a trigger-action rule is ``IF I am not at home, THEN turn on the camera''.  A trigger can be a cyber or physical event reported to the smart home system by the IoT service, such as a motion-sensing event. An action is a physical change like turning on a light.  A variety of TAP  platforms, such as IFTTT\footnote{\url{https://ifttt.com/}}, Zapier\footnote{\url{https://zapier.com/}}, SmartThings\footnote{\url{https://smartthings.com/}}, and openHAB\footnote{\url{https://openhab.org/}}, empower non-technical occupants to compose IoT services and Web services (e.g., social media and message apps \cite{corno2020taprec}).

With an increasing number of IoT rules created by occupants, inter-rule vulnerabilities can be introduced accidentally. An inter-rule vulnerability refers to a situation resulting from the interaction between IoT services. We identify two broad categories of inter-rule vulnerabilities: \emph{explicit inter-rule vulnerability} and \emph{implicit/physical inter-rule vulnerability}. An explicit inter-rule vulnerability is manifested when two IoT rules compete to act on the shared IoT services, e.g., turning on and turning off the same light are requested by two IoT rules simultaneously. An implicit inter-rule vulnerability (a.k.a \emph{physical inter-rule vulnerability}) refers to the situation that two IoT rules interact with each other through a shared environment channel/property, such as temperature, humidity, and illumination. One feature of IoT services is their capability of interacting with physical environments such as temperature, humidity, and illumination. For instance,  a heater service can increase the room temperature. As a result, the physical interaction capabilities enable multiple IoT rules to interact with each other through a shared environment.  The physical inter-rule vulnerability may cause a series of unexpected consequences ranging from a less comfortable home environment to security issues. For example, an IoT rule requests to turn on the air-conditioner to cool the room while another rule requests to open the window in hot summer \cite{hua2022copi}. Opening the window perhaps lets hot air come in and reduces the air-conditioner's cooling effect. Furthermore, the physical inter-rule vulnerability may be exploited by attackers to achieve malicious purposes. An IoT rule that changes physical environments may trigger the execution of another IoT rule accidentally via their shared environment channels. For example, a \emph{Heating} service turns on the heater when the room temperature is less than 20$^{\circ}$C, while the \emph{Cooling} service opens the window when the temperature is more than 25$^{\circ}$C (i.e., temperature $<$ 20$^{\circ}$C $\rightarrow$ turn on the heater, temperature $>$ 25$^{\circ}$C $\rightarrow$ open the window). Suppose that an attacker has obtained access to the heater service, he/she can maliciously turn on the heater to gradually increase room temperature thereby executing the \emph{Cooling} service to open the window.

%\indent 

Current works mainly focus on identifying explicit inter-rule vulnerabilities from IoT app source code \cite{yu2022tapinspector,chi2020cross,alhanahnah2020scalable,chen2019multi}. An IoT app can be modeled as a trigger-action rule \cite{chi2020cross}. These works first translate IoT apps into trigger-action rules by performing static code analysis. Then, explicit inter-rule vulnerabilities are identified by comparing the device's attributes such as turning on and turning off the lights requested in different rules. There are a few studies that consider physical inter-rule vulnerabilities \cite{ding2018safety,ding2021iotsafe}. They rely on static code analysis to extract trigger-action rule information from app source code and use NLP techniques to extract environment channels from rule descriptions. 
These works adopt a \emph{posteriori}  approach in that physical inter-rule vulnerabilities are detected after the IoT rule is deployed or the IoT system runs. In this paper, we adopt an \emph{apriori} approach that \emph{proactively} detects physical inter-rule vulnerabilities from user requirement specifications (i.e., descriptions) at the IoT rule design stage and prevents  them from being introduced into the smart home system. Inspired by defect inspection in software development, the earlier the detection of requirements defects, the easier it will be to correct them at a lower cost \cite{alshazly2014detecting,huang2017software}.  To this end, we propose a  deep learning-based approach for automating extracting rules from their descriptions and detecting physical inter-rule vulnerabilities hidden among these extracted rules. In particular,   we first extract rules from descriptions specified in natural language. For each extracted rule,  we identify environment channels using the latest NLP tools. Finally, given the extracted rules and their associated environment channels, we identify possible physical inter-rule vulnerabilities. In a nutshell, the main contributions of our work are as follows: 
\begin{itemize}
\item[$\bullet$] We propose a novel framework that extracts IoT rules from requirement specifications and identifies potential physical inter-rule vulnerabilities, intending to avoid introducing them into smart home systems. Compared to existing works that detect and eliminate vulnerabilities \emph{after} the IoT rule deployment,  to the best of our knowledge, our work is the \emph{first} attempt to proactively discover physical inter-rule vulnerabilities from requirements at the IoT system design stage to prevent them from being introduced to the system.
\item[$\bullet$] We propose a  novel approach for extracting IoT rules specified in natural languages. It adapts the state-of-the-art   Transformer sequence-to-sequence (seq2seq)architecture to translate an input requirement into an IoT rule.  
\item[$\bullet$] For each extracted IoT rule, we associated it with environment channels using the latest NLP tools.  We formalize two types of physical inter-rule vulnerabilities based on the trigger-action rule model and environment channels.  

\item[$\bullet$] We conduct extensive experiments on two real-world datasets including IFTTT rules and SmartThings IoT apps.   Our experimental results on IFTTT rules show that the proposed Transformer approach can effectively extract rules with 95.22\% accuracy. Our physical inter-rule vulnerability detection approach can detect 99 possible vulnerabilities hidden among 60 IoT apps.  
\end{itemize}
The paper is organized as follows. Section 2 provides an overview of the framework.  Sections 3, 4, and 5  describe the IoT rule extraction,  the environmental channel identification, and physical inter-rule vulnerability detection modules, respectively.  Section 6 presents the experimental results. Related work is presented in Section 7 and finally, we conclude in Section 8.

\section{Framework Overview}
\begin{figure*}[htbp]
 \vspace{-2mm}
\centering
\includegraphics[width= 0.7\textwidth, height=0.25\textwidth]{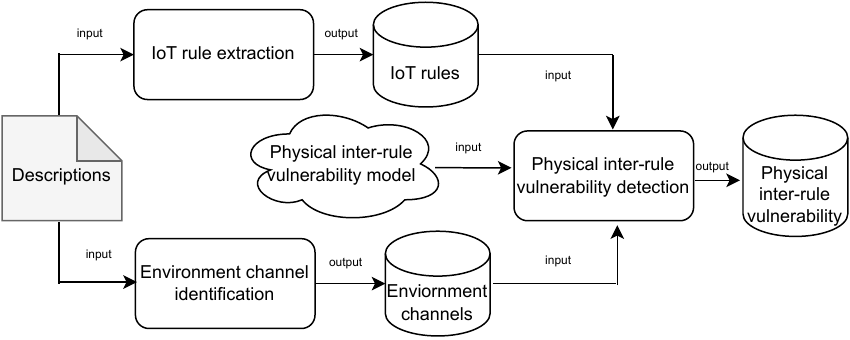}
\caption{Overview of the proposed framework}
\label{fig:frameworkoverview}
\vspace{-3mm}
\end{figure*}

This section provides an overview of the framework in Fig.\ref{fig:frameworkoverview}.  Our framework constitutes three modules including IoT rule extraction, environment channel identification, and physical inter-rule vulnerability detection. The IoT rule extraction module in Section \ref{method} uses the Transformer to extract trigger-action rules (i.e., represented by a sequential pair of IoT services) from a set of descriptions. Given an extracted IoT rule and its associated description, the environment channel identification module in Section \ref{physical-channel-identification} is responsible for finding environment channels from this description. Taking a set of extracted rules and their associated environment channels as inputs, the physical inter-rule vulnerability detection module in Section \ref{physical-inter-rule detection} focuses on discovering possible vulnerabilities. 

\section{IoT Rule Extraction}\label{method}
In this section, we first propose a trigger-action rule model and present a deep learning approach for extracting trigger-action rules from natural language descriptions.

\subsection{Trigger-Action Rule Model}\label{method:trigger-action-rule-model}  
We first formalize an IoT service based on functionalities and non-functional properties (e.g., environment channels).  Then, we define the trigger-action rule based on the IoT service model.  

\vspace{2mm}
\noindent \textbf{IoT Service}. An IoT service is  described by a tuple $s = < id, \alpha, eff > $ where:
\begin{itemize}
\item[$\bullet$] $id$ is a unique identifier for the smart object. 
  
\item[$\bullet$] $\alpha = \{\alpha_1, \alpha_2,..., \alpha_n \}$ is a set of functionalities or methods offered by the service. 
 
\item[$\bullet$] $eff = \{ env_1, env_2,.., env_n\}$ is a set of \emph{environment channels/properties} associated with the service. An environment property $env_i$ is impacted if the usage of the service $s$ changes the environment value of $env_i$.  
\end{itemize}
For example, a fan service offers functionalities including turning on/off, and setting fan speed and has the capability of changing temperature. It can be described as $\langle$ \{fan.on,  fan.off,  fan.setFanSpeed  \}, \{temperature\}$\rangle$. We use environment channels and environment properties interchangeably. 

\vspace{2mm}
\noindent \textbf{Trigger-Action Rule}. An IoT rule or trigger-action rule $r$ is represented by $trig\rightarrow f$ where:
\begin{itemize}
\item[$\bullet$] $trig = \langle trigger\_title, trigger\_channel \rangle $ is the trigger component where $trigg\-er\_title$ is the trigger service name,  $trigger\_channel$ is a service event type which can be an IoT service state (i.e., ON and OFF states of a light service), environment properties (i.e.,  temperature), time, and the occupant's presence.   
\item[$\bullet$] $f = \langle action\_title, action\_channel \rangle $ is the action component where $action\_title$ is a service name, $action\_channel$ is a service functionality.
\end{itemize}
For example, the rule ``$\langle$thermometer, temperature  $\rightarrow$ AC,  AC.on $\rangle$'' means turning on the air-conditioner when the temperature reaches a certain value.  It should be noticed that time and the resident's presence can also be triggers. We do not consider them as they play little role in causing physical inter-rule vulnerabilities. 
\subsection{A Deep Learning Approach for IoT Rule Extraction}\label{method:iot-rule-extraction}
This section presents a deep learning approach for extracting IoT rules from descriptions. It is challenging to extract four elements  $\langle trigger\_title, trigger\_channel \rangle $  $\rightarrow $ $ \langle action\_title, \linebreak action\_channel \rangle $ as a whole and ensure their correct positions. Our target is to find environment channel interactions between IoT rules. An alternative method is to identify environment channels for each IoT service from outsourced knowledge or human annotations.  For example, the knowledge graph ConceptNet is used to identify environmental channels \cite{huang2021conflict}. Therefore, we address this challenge by simplifying the IoT rule into  $\langle trigger\_title  \rangle $  $\rightarrow $ $ \langle action\_title \rangle $ and identifying environment channels from outsourced knowledge using NLP tools. In the rest of the paper, an IoT rule is simplified as a sequential pair of IoT services in the form of  $\langle trigger\_title  \rangle $  $\rightarrow $ $ \langle action\_title \rangle $.

\begin{figure*}[htbp]
 \vspace{-2mm}
\centering
\includegraphics[width= 0.65\textwidth, height=0.23\textwidth]{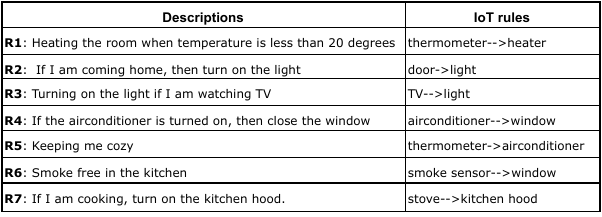}
\caption{Extracting IoT rules from descriptions}
\label{fig:graph}
\vspace{-3mm}
\end{figure*}

Each IoT rule is associated with a description where an occupant specifies the expected functionality of the rule using natural language. Examples of descriptions are shown in Fig. \ref{fig:graph}. Several challenges are posed in generating trigger-action rules from descriptions: \textbf{1) Vague description}. We identify two categories of descriptions: specific and vague descriptions.   A  description is considered specific if it is described by \emph{explicitly} referring to IoT services.  For example, the requirement ``If the air-conditioner is turned on, then close the window''   explicitly shows the rule ``air-conditioner $\to$ window''. A description is considered vague if it is described using high-level and general words that \emph{implicitly} refer to particular IoT services. For example, a vague description is ``If I am coming home, then turn on the light''. It implies the rule ``door $\to$ light''. Users are prone to express their high-level intention using vague descriptions \cite{corno2020taprec}. Studies show that 80\% of the descriptions of  IFTTT rules (a.k.a., recipes) are ambiguous because they do not explicitly mention the involved services. We find similar results in a recent IFTTT dataset used in our study that 82.8\% of descriptions are vague. \textbf{2) Uncertain ordering}.  The trigger and the action may appear in different orders in two different descriptions that express the same meaning. For example, ``If I am watching TV, then turn on the light'' (i.e., TV's position is before that of the light) and ``Turning on the light when I am watching TV'' ((i.e., TV's position is after that of the light)) has a similar meaning. However, the ordering of the light and TV are opposite in the two similar descriptions. \textbf{3) Varied length}. The length of descriptions is varied for different rules.

To address the aforementioned challenges, we use the Transformer sequence-to-sequence model  to generate trigger-action rules from descriptions.  Formally, given a set of requirement descriptions for rules 
$$Req = \{ (req_1, r_1), (req_2, r_2) \cdots (req_n, r_n)\}$$
where $(req_i, r_i)$ denotes a description and its associated trigger-action rule (i.e., (``If the AC is turned on, then close the window.'', ``AC $\rightarrow$ window'')), we aim to train a Seq2Seq model that takes a $req_i$ as input and generates the trigger tile and action tile to form the rule $r_i$. 

Transformer models, which are based on the attention mechanism \cite{vaswani2017attention}, are commonly pre-trained on massive corpus and excel in a range of natural language processing tasks. Our approach fine-tunes a pre-trained Transformer Seq2Seq model to extract the trigger-action rules from plain text-based descriptions. 

\begin{figure}[!ht]
    \centering
    \includegraphics[width=0.95\linewidth, height=0.45\textwidth]{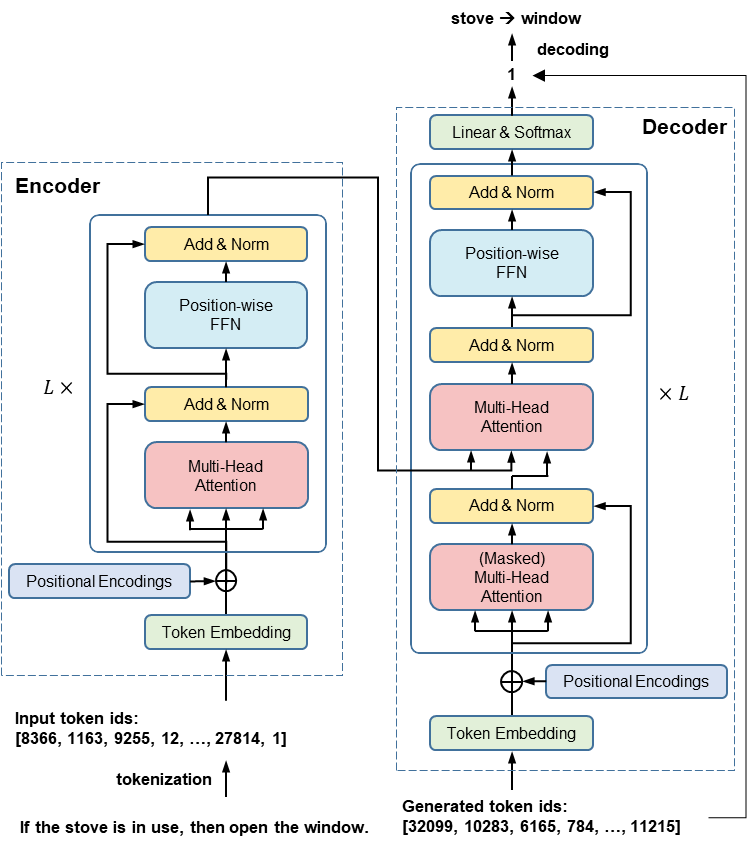}
    \caption{The architecture of the Transformer model.}
    \label{fig:transformer}
    \vspace{-3mm}
\end{figure}

As shown in Fig. \ref{fig:transformer}, the Transformer has the \emph{encoder} and the \emph{decoder} components, each 
%of which 
constituting a stack of $L$ identical attention blocks. The process of generating triggers and actions is illustrated in Algorithm~\ref{transformer}.
For each input description  $req_i$ in $Req$, it is first tokenized into a sequence of token ids via a pre-trained tokenizer which records mappings between text fragments and token ids:
\begin{align}
\mathbf{T}_{in}=tokenizer(req_i)
\end{align}
Second, an embedding layer converts the tokens $\mathbf{T}_{in}$ into token embedding $emb_e$ with a dimension of $d_{model}$ (i.e., the predefined input dimension to the model):
\begin{align}
\label{emb}
emb_e=Embedding(\mathbf{T}_{in})
\end{align}
Third, we calculate the positional encodings~\cite{vaswani2017attention} to inject the position information of the tokens in the text,
\begin{align}
\label{equ:pe1}
PE_{(pos, 2i)}=sin(pos/10000^{2i/d_{model}})\\
\label{equ:pe2}
PE_{(pos, 2i+1)}=cos(pos/10000^{2i/d_{model}})
\end{align}
where $pos$ is the position and $i$ is the dimension.

\noindent Fourth, the \emph{Encoder} maps the summation of $emb_e$ and $PE_e$ to an encoded vector $E$:
\begin{align}
E=Encoder(emb_e+PE_e)
\end{align}
Fifth, the \emph{Decoder} auto-regressively produces the output tokens with the $E$ and the decoded tokens from the previous decoding step until an \textless EOS\textgreater~(End of Sequence) token is produced. The prediction of the next token is shown as follows:
% TODO: add softmax to predict the most probable output token
\begin{align}
\label{equ:dvec}
D=Decoder(E, Embedding(t^{j}_{out})+PE_d)\\
\label{equ:tout}
t^{j+1}_{out}=\mathrm{argmax}(\mathrm{softmax}(D))
\end{align}
where $t^0_{out}$ equals to \textless BOS\textgreater~(Begin of Sequence).
%Last, 
Finally, the tokenizer converts the decoded tokens back into words as the final output.  %We elaborate on the Transformer architecture as follows. 

\begin{algorithm}[!tb]
\caption{Transformer for Generating Trigger-action Rules}
\label{transformer}
\scriptsize
 \begin{algorithmic}[1]
 \renewcommand{\algorithmicrequire}{\textbf{Input:}}
 \renewcommand{\algorithmicensure}{\textbf{Output:}}
  \REQUIRE  $Req$ (a set of descriptions); 
  %transformer $Encoder$ and $Decoder$; $W_l, b_l$ linear classifier weight and bias.
  \ENSURE  $R$ (a set of extracted IoT rules)
  % \STATE Set $R'$$ as 
\FORALL{$req_i$ in $Req$}
  \STATE Tokenize and extract embedding $req_i$ as $emb_e$  
  \STATE Get the positional encoding $PE_{e}$ by formula (\ref{equ:pe1}) and (\ref{equ:pe2})
  % \STATE $emb_e \leftarrow emb_e + PE_{pos}$
   % \FOR{$l = 1\ to\ L$} 
   %       \STATE $E^{(l+1)}$ $\leftarrow$ $Encoder(E^{(l)})$
   % \ENDFOR
   \STATE Extract the encoding vector via the Encoder as $E$ by formula (\ref{emb})
   
   \STATE $t^0 \leftarrow $ [BOS]   %\\ initialize $gen$ with the predefined begin of sequence token
   \STATE $j \leftarrow 1$
   \WHILE{$t^{j-1}$ is not [EOS]}
        \STATE Tokenize $t^{j-1}$ and extract the embedding as $emb^j_d$
        \STATE Get the positional encoding $PE^j_{d}$ by formula (\ref{equ:pe1}) and (\ref{equ:pe2})
        % \STATE $emb_d \leftarrow emb_d + PE_{d}$
        \STATE Decode and predict the next token $t^j$ by formula (\ref{equ:dvec}) and (\ref{equ:tout})
        \STATE $j \leftarrow j+1$
        % \FOR{$l = 1\ to\ L$} 
        %     \STATE $D^{(l+1)}$ $\leftarrow$ $Decoder(E^{(L)}, D^{(l)})$
        % \ENDFOR
        % \STATE $next \leftarrow \mathrm{argmax}(\mathrm{softmax}(FFN(D))$
        \STATE $T_{out}.append(t^j)$
   \ENDWHILE
   
    \STATE Convert all tokens in $T_{out}$ to texts via the tokenizer and add them to $R$
\ENDFOR
\RETURN{$R$}
\end{algorithmic}
\end{algorithm}

\section{Environment Channel Identification}\label{physical-channel-identification}
Environment channel identification (i.e., physical channel identification) aims at extracting environment properties from descriptions.  An environment channel is an environment entity through which IoT services interact with each other indirectly. We consider six common environment entities including \emph{temperature, illumination, smoke, sound, humidity and air quality}. We employ the latest ChatGPT-4 to identify environment channels from descriptions. ChatGPT-4 has shown great potential in processing text data in the task of entity recognition \cite{hu2023zero}\cite{wei2023zero}. The key task of asking  ChatGPT-4 is to design the proper prompts. We directly input descriptions into 
%the 
ChatGPT-4 and instruct it to extract environment channels.

\vspace{2mm}
\noindent \textbf{Prompt Instruction}:  \emph{``Please compute the similarity between a sample in the sentence list with every element in the keyword list. The output should be the most similar keyword with a similarity score for each sample. If no keyword is matched, print ``None''. The output is in the format of: ``keyword, score'' or ``None''.  The keyword list is [``temperature'',``illumination'',``humidity'',``smoke'',``sound'' ,``air quality'']. The sentence list  is   [$des_1$, $des_2$...$des_n$ ].''}

\vspace{2mm}
\noindent\textbf{Ground Truth and Performance}: We collect descriptions of 60 IoT apps and manually annotate each involved IoT service with environment channels as ground truth. For example, light is associated with ``illumination''. Some devices, such as doors and windows, may have multiple environment channels in different smart home layouts. The context of doors and windows greatly influences their environment channels. We conduct two groups of physical channel identification. In the first group, all IoT services, except for doors and windows, are appropriately assigned environment channels. In the second group, all devices are assigned proper environment channels in which doors and windows are assigned with \emph{temperature}, \emph{humidity}, and \emph{illumination}. 

The outputs are collected from the online ChatGPT-4 interface and compared with the ground truth. We calculate the accuracy of correctly identified environment channels. In the first group, the accuracy of physical channel identification is 93.3\%, whereas it is 83.3\% in the second group. The results show  that ChatGPT-4 performs well in identifying the main effects of an IoT service on the environment, whereas performs poorly in inferring an IoT service's side effects on the environment, e.g., 
the primary environmental effect of the air-conditioner is reducing the temperature, while its secondary environmental effects include generating noise and reducing humidity. 

\section{Physical Inter-rule Vulnerability Detection}\label{physical-inter-rule detection}
This section first formally defines two types of physical inter-rule vulnerabilities (i.e., Environment Channel Rule Chain  and  Environment Channel Interference) based on the trigger-action rule model and environment channels.  Then it describes the process of the physical inter-rule vulnerability detection module. 

\vspace{2mm}
\noindent \textbf{Environment Channel Rule Chain:} 
An Environment Channel Rule Chain exists if the environmental impact of executing an IoT rule triggers or disables the execution of another rule accidentally. For example, a rule turns on the humidifier to make the air more humid in the living room. The air humidity increases after turning on the humidifier for a while. As a result, the increasing air humidity may activate the dehumidifier automatically to dehumidify the air. Formally, given two IoT rules $r_i$ and  $r_j$, the action title service $s_i$ in the former rule has a set of environment channels $eff_i = \{ env_1, env_2,.., env_n\}$. These environmental channels may be the triggering channel of the latter rule. The following condition is satisfied:
\begin{equation}
eff_i \cap  trigger\_channel^j \neq  \phi
\end{equation}

%\vspace{2mm}
\noindent \textbf{Environment Channel Interference:} 
An Environment Channel Interference refers to the situation that two rules update the shared environment properties. For example, a rule requests to use the air-conditioning to cool the living room in hot summer while another rule opens the window to let outside fresh air come in. An environment channel interference may occur because opening the window would let hot air come in and impact the air-conditioning's cooling effect.  Formally, given two IoT rules $trig^i\rightarrow f^i$ and  $trig^j\rightarrow f^j$, their action titles are the service $s_i$ and $s_j$, respectively. The environment channels associated with   $s_i$ and $s_j$ are $eff_i = \{ env_1, env_2,.., env_n\}$ and $eff_j = \{ env_1, env_2,.., env_m\}$, respectively. The following conditions should hold:
\begin{equation}
   (s_i\neq s_j)  \wedge (eff_i \cap   eff_j \neq  \phi)
\end{equation} 
 
The physical inter-rule vulnerability detection module takes all extracted trigger-action rules and identified environment channels as inputs. The outputs are all possible vulnerabilities (i.e.,  Environment Channel Rule Chains or   Environment Channel Interference), which are generated by connecting inter-rule interactions through proper environment channels. 

Specifically, this module detects physical inter-rule vulnerabilities from a set of extracted rules in a \emph{particular} smart home setting. In a particular smart home setting, we assume the \emph{location} information of each IoT service is available because IoT services in different locations are less likely to be physically interacting. In this regard, physical inter-rule vulnerability detection is conducted for IoT services that are located at the same location. In this work, we collect and assign a semantic location such as a bathroom and bedroom to each IoT service.   We use a dictionary to store rules. 
Given two rules $r_i: trig^i\rightarrow f^i$ and $r_j: trig^j\rightarrow f^j$, we check whether there are vulnerabilities between them based on the physical inter-rule vulnerability model defined in Equation (8) and (9). 

\vspace{-2mm}
\section{Experimental Results}\label{experiments}
\vspace{-2mm}
We conduct a  set of experiments to evaluate   our proposed framework. The first set examines the performance of extracting trigger-actions rules from descriptions.    The second set evaluates the effectiveness of detecting physical inter-rule vulnerabilities.   
The experiments are implemented in Python and use open-source packages Pytorch\footnote{\url{https://pytorch.org/}} 
and HuggingFace \cite{wolf2020transformers}. The experiments are performed on an Ubuntu 20.04 server with AMD EPYC 7763 CPU@2.45GHz, NVIDIA RTX A6000 (49GB).

\vspace{2mm}
\noindent \textbf{Experiment I: Performance of the Transformer in extracting trigger-action rules from descriptions}. 
This experiment aims to evaluate the performance of the Transformer in generating trigger-action rules from descriptions. It is conducted on the latest IFTTT rules\footnote{\url{https://www-users.cse.umn.edu/~fengqian/ifttt_measurement/}} (a.k.a., recipes) collected  from the popular trigger-action programming  platform IFTTT in May 2017. It contains 27983 rules.  
We preprocess the raw dataset by removing noise data that are not written in English. We join the \textit{Tile} and \textit{Description}   columns as the   description and use the \textit{(triggerTitle, actionTitle)} columns as the ground truth value i.e., the sequential IoT service pairs.   An example of the dataset is $\langle$(``Turn off lights when motion is no longer detected by Wyze Motion Sensor''), (Wyze, Philips Hue)$\rangle$. We use $accuracy$ as the metric of the performance, which is defined as follows: 
 \begin{equation}
   accuracy   = \frac {|correct|}{|samples|}
\end{equation}
where $|samples|$ is the total number of samples and $|correct|$ is the number of correctly predicted trigger tile and action title that are aligned sequentially. It should be noted that a sample counted as \emph{correct} must meet two constraints at the same time: 1) both the predicted trigger title and action title must exist in the ground truth sample, and 2) the sequential relation between the predicted trigger title and action title must be the same as the one in the corresponding ground truth sample.  

 \begin{table}[!htbp]
 % \vspace{-6mm}
\caption{Hyperparameter settings for the Transformer}
\centering
	\begin{tabular}{lccr}
\toprule
 Hyperparameters &Values \\
\midrule

  batch size($\mathcal{B}$) & 32 \\  
           learning rate ($\eta$) & 5e-4  \\  
           number of beams ($n$) & 5  \\  
           epoch ($ep$) & 50 \\
           optimizer &  AdamW \\ 
           search algorithm & Beam search\\

\bottomrule
\end{tabular}
\label{tab:Hyperparameter settings for the Transformer}
%\vspace{-0.7cm}
\end{table}

We utilize T5~\cite{T5} as our backbone model due to its effectiveness on diverse NLP tasks. Specifically, we initialize our model with T5-large and follow the default settings of T5 to fine-tune our Transformer model as shown in Table \ref{tab:Hyperparameter settings for the Transformer}. We set batch size $\mathcal{B}$, learning rate $\eta$, number of beams $n$ and epoch $ep$ to be 32, 5e-4, 5, and 50, respectively. We choose the AdamW optimizer for the Gradient descent algorithm and use the standard beam search decoding algorithm~\cite{Seq2Seq} to find the best prediction results. We use 80\% and 20\% of the dataset  for training and testing, respectively. 
%TABLE
Table \ref{tab:transformer result} shows the accuracy of the Transformer in generating trigger-action rules.  The Transformer achieves an overall accuracy score of approximately 95.22\% on the test samples. 

 \begin{table}[!htbp]
 \vspace{-6mm}
\caption{ Results of the Transformer in generating trigger-action rules.}
\centering
	\begin{tabular}{lccr}
\toprule
 Groups &\# Samples &$accuracy_T$\\
 
\midrule
Specific descriptions & 4799 & 99.29\% \\  
Vague descriptions & 23184 & 94.37\%\\  
All & 27983 &95.22\% \\ 
\bottomrule
\end{tabular}
\label{tab:transformer result}
%\vspace{-0.5cm}
\end{table}

A key feature of the dataset is that it contains both \emph{specific} and \emph{vague} descriptions. A description is considered specific if it is described by \emph{explicitly} referring to   IoT services.  For example, the description ``If the air-conditioner is turned on, then close the window''   explicitly shows the rule ``air-conditioner $\to$ window''. A description is considered vague if it is described using high-level and general words that \emph{implicitly} refer to particular IoT services. For example, a vague description is ``If I am coming home, then turn on the light''. It implies the rule ``door $\to$ light''. Therefore, the Transformer model should be capable of learning specific IoT services from vague descriptions. To study such generalization capability of the Transformer, we split the dataset into two groups: specific description group containing  4,799 samples, and vague descriptions containing 23,184 samples. We test the Transformer on these two groups. As expected, the Transformer achieves much higher accuracy (99.29\%) in the specific description group than that in the vague description group. The experiment result demonstrates that the Transformer performs well in extracting trigger-action rules from descriptions and generalizing well on vague descriptions that express a high level of user intention.

\vspace{2mm}
\noindent \textbf{Experiment II: Effectiveness of physical inter-rule vulnerability detection}. 
This experiment aims to evaluate the effectiveness of our physical inter-rule vulnerability detection approach. We study 185 official IoT apps provided by the SmartThings platform\footnote{https://github.com/SmartThingsCommunity/SmartThingsPublic}. 
Some IoT apps are too complex that are implementations of multiple trigger-action rules, which would generate excessive physical inter-rule vulnerabilities. Some IoT apps only use  Web services that don't interact with physical environments. We exclude these two types of IoT apps using the SmartVisual tool \cite{bak2020smartvisual} and finally use 60 representative IoT apps for our evaluation.

Each IoT app is associated with a description. For example, the app \emph{brighten-my-path} has the description 
``Turn your lights on when motion is detected''. Given a description of an IoT app,  we manually extract its trigger-action rules. Next, we identify environment channels for each trigger-action rule. We use the latest ChatGPT-4 to identify environment channels as described in 
%michael: it is not good to hard code things. Pls double check and ensure the section number is correct. 
Section \ref{physical-channel-identification}.  
It successfully identifies three environment channels in total including \emph{temperature, humidity}, and \emph{illumination}.  Our environment channel identification analysis successfully connects 93.3\% (56 out of 60 IoT apps) of the IoT rules with the correct environment channels. The failures mainly come from the case that a description contains no environment channel-related information. For example, the description ``Turn on one or more switches at a specified time and turn them off at a later time'' is not assigned with any environment channels even if the switch capability is subscribed by lights.

We compare each pair of IoT rules and check possible physical inter-rule vulnerabilities using 
%michael: not good to hard code. Pls double check and ensure they are correct. 
Equations (8) and (9), and  
discover 99 physical inter-rule vulnerabilities. 60.6\% (60 out of 99) of them are Environment Channel Rule Chain vulnerabilities that one rule execution may trigger or disable the execution of another rule via the identified environment channel (i.e., temperature, humidity, and illumination). The rest are Environment Channel Interference vulnerabilities that two rules may update the shared environment channel simultaneously.

\textbf{Discussion}: We propose a deep learning-based approach (Transformer) to extract trigger-action rules from descriptions and discover hidden physical inter-rule vulnerabilities via shared environment channels. The noticeable novelty of our approach is that it proactively identifies physical inter-rule vulnerabilities at the IoT rule specification stage, thereby preventing introducing them into the smart home system. However, several limitations deserve attention: (1). The accuracy of environment channel identification highly relies on the context of a given smart home. For example, the window in one smart home impacts temperature while it may not have such a feature in another home. External information,  such as the layout of a home, or occupants' feedback can help in improving the accuracy of environment channel identification. 
(2). Some physical inter-rule vulnerabilities are severe and risky and need to be resolved immediately, while others are less serious and can be ignored. An assessment mechanism is needed to evaluate the risk levels of the vulnerabilities.

\section{Related Work}\label{related-work}
Our work lies at the intersection of research in three related areas: trigger-action programming, rule extraction from natural language descriptions, and inter-rule vulnerability detection. 

\subsection{Trigger-action programming}\label{related-work:trigger-action-programming}
The proliferation of IoT services brings many opportunities to smart homes, making residents' home lives more convenient, efficient, secure, and entertaining by automating IoT services \cite{huang2021enabling}. End-User Development (EUD) focuses on providing technologies and tools and putting the IoT-based applications/services development in the hands of residents who are most familiar with the actual needs but have limited programming skills \cite{ghiani2017personalization}. 
%End-user development (EUD)  
EUD is a promising approach to automate and remotely control IoT services, which empowers residents who have little programming knowledge to customize smart homes by trigger-action programming. A trigger-action program can be expressed by IF-THEN rules in the form of ``IF triggers
happen, THEN perform an action'' \cite{corno2019recrules}. 
%michael: no need to give an example here. We already gave lots of examples. 
%An example of a trigger-action rule is ``IF I am not at home, THEN turn on the video camera.''.  
Triggers and actions are connected by conditions that are logical
predicates defined by the current state of the triggering events. If the triggering event makes the condition true, the corresponding action is exerted.  
%michael: exact sentences already in the introduction. Suggest to simplify this sentence.
%A variety of EUD platforms, such as IFTTT\footnote{\url{https://ifttt.com/}}, Zapier\footnote{\url{https://zapier.com/}}, SmartThings\footnote{\url{https://smartthings.com/}}, and openHAB\footnote{\url{https://openhab.org/}}, empower non-technical residents to compose IoT services and Web services (e.g., social media and message apps \cite{corno2020taprec}). 
Several EUD platforms, e.g., IFTTT, Zapier, SmartThings, and OpenHAB, offer the tools so that non-technical residents can compose IoT services. 

%michael: too big paragraph. Split into a few
\subsection{Inter-Rule Vulnerability Detection}\label{related-work:detection}
Inter-rule vulnerability is becoming a critical issue in trigger-action programming platforms. It is estimated that around 15\% of rules on the IFTTT platform have inter-rule vulnerabilities \cite{wu2020learning}. 
The majority of studies perform \emph{static code analysis} on IoT app sources to extract TAP rules and then inspect inter-rule vulnerabilities \cite{yu2022tapinspector,chi2020cross,alhanahnah2020scalable,chen2019multi,ding2018safety,ding2021iotsafe,ozmen2022discovering,wang2019charting}. Some of these works consider physical inter-rule vulnerabilities \cite{ding2018safety,ding2021iotsafe,ozmen2022discovering,wang2019charting}. 

The proposed framework, \emph{IoTMon}, aims to discover physical interactions between IoT apps \cite{ding2018safety}. These interactions occur when IoT devices communicate through shared environment channels like temperature, humidity, and air. The framework extracts information from the app's source code to create intra-app interactions and employs NLP techniques to identify physical channels from the app's descriptions. By chaining together these interactions through the identified physical channels, the framework generates interaction chains. The risk level of these interaction chains is evaluated based on the similarity score between intra-app interactions of trustworthy applications. 
In \cite{ding2021iotsafe}, the proposed system, \emph{IoTSAFE}, focuses on detecting physical interactions between IoT devices in smart homes. It utilizes a run-time physical interaction discovery approach that involves static code analysis to construct an interaction graph and dynamic testing techniques to discover real-time physical interactions among IoT devices. The system also leverages contextual features to predict future risky situations and disable unsafe device states using the interaction graph and the temporal physical interaction graph. 

The proposed framework, \emph{IoTSeer}, is designed to identify physical interactions between IoT apps \cite{ozmen2022discovering}. These interactions involve IoT devices communicating through shared environment channels.
%like temperature, humidity, and air. 
The framework extracts information from the app's source code to create intra-app interactions and utilizes NLP techniques to identify physical channels from the app's descriptions. By linking these interactions through the identified physical channels, the framework generates interaction chains. The risk level of these chains is evaluated by comparing their similarity to intra-app interactions of trustworthy applications. A system called \emph{iRULER} is developed to identify inter-rule vulnerabilities that exist within trigger-action platforms \cite{wang2019charting}. It identifies six types of vulnerabilities and formally defines them. It extracts inter-rule information flows by using Natural Language Processing (NLP) to inspect the text descriptions of triggers and actions on the IoT platform website. It performs Satisfiability Modulo Theories (SMT) solving and model checking to discover inter-rule vulnerabilities using extracted information flow.  

%michael: too big paragraph. Split into a few
\subsection{Trigger-Action Rule Generation}\label{related-work:trigger-action-rule-generation}
The task of trigger-action rule generation is to predict elements including trigger title, trigger channel, action tile, and action channel from descriptions. Prior works treat the problem of trigger-action rule generation as a multi-task classification task. A proposed approach to address this problem involves combining a neural network with logistic regression \cite{beltagy2016improved}. It formulates the problem of natural language understanding as a structure prediction problem. The predicted element is conditioned on the whole input and all prior decisions. The existing work proposes a Latent Attention Model (LAM) to generate trigger actions automatically \cite{liu2016latent}. Taking a natural language description of the intended TAP as the input, LAM identifies the contained channel and function for trigger and action respectively. For example, given the description ``Autosave your Instagram photos to Dropbox'', LAM predicts its trigger channel ``Instagram'' has the trigger channel Instagram, trigger function ``Any new photo by you'', action channel ``Dropbox'', and action function ``Add a file from URL''. LAM utilizes a latent attention mechanism to locate the words in the description that are the most relevant for predicting desired labels. 

The current state-of-the-art research frames the TAP generation as a sequence-to-sequence learning problem \cite{dalal2020evaluating}. A recent study proposes an approach RecipeGen to automatically generate TAP rules given natural language descriptions \cite{yusuf2022accurate}. RecipeGen treats the TAP generation as a natural language translation problem. It leverages Transformer sequence-to-sequence (seq2seq) architecture to generate a sequence of triggers and actions from natural language descriptions. Compared to LAM which performs poorly when the description is vague,  RecipeGen generalizes well and performs better because it learns implicit relations between the channels, functions, and fields of triggers and actions. A Hierarchical Reinforcement Learning (HRL) approach is proposed to parse trigger channels, trigger functions, action channels, and action functions from a natural language description \cite{yao2019interactive}. It decomposes the task into 4 subtasks (i.e., predicting trigger/action channel/function), and 
%.  It 
designs hierarchical policies: a high-level policy on deciding the order of the task and a low-level policy on deciding whether to ask users a clarification question. The policies are trained to maximize the parsing accuracy and minimize the number of questions with the rewarding mechanism.

\section{Conclusion and Future Work}\label{conclusion}
In this paper, we propose a novel approach to proactively discover and prevent physical inter-rule vulnerabilities of Internet of Things (IoT) services in smart homes. We particularly rely on a deep learning model (Transformer) to accurately learn trigger-action rules from natural language descriptions that are associated with the rules. We also utilize the latest ChatGPT 4 to identify environment channels for each extracted rule with high accuracy. We propose two types of physical inter-rule vulnerabilities and formalize them based on the formal trigger-action rule model. The experiments on real-world datasets demonstrate the high performance of the Transformer in generating trigger-action rules. We also validate the effectiveness of the approach in detecting physical inter-rule vulnerabilities on existing IoT apps. In the future, we plan to devise a risk-level assessment mechanism to evaluate the severity of physical inter-rule vulnerabilities. We also plan to learn occupants' preferences towards environmental properties for dynamically resolving physical inter-rule vulnerabilities. 

\section*{Acknowledgment}
The research for this project is supported by Cyber Security Agency of Singapore (CSA).

\bibliographystyle{IEEEtran}
\bibliography{references}
\end{document}